\documentclass{PoS}

\usepackage{cite}
\usepackage{tabularx}
\usepackage{amsmath}
\usepackage{float}
\usepackage{datetime}
\usepackage{textpos}
\usepackage{booktabs}
\usepackage{graphicx}
\usepackage{mathrsfs}

\providecommand{\abs}[1]{\lvert#1\rvert}

\title{Can Cosmic Rays Interacting With Molecular Clouds Explain The Galactic Center Gamma-Ray Excess?}

\ShortTitle{Galactic Center Gamma-Ray Excess}

\author{\speaker{Chris Gordon} and Oscar Macias
\\
        University of Canterbury, New Zealand\\
        E-mail: \email{chris.gordon@canterbury.ac.nz}}

\abstract{The Fermi Large Area Telescope data  appear to have an excess of gamma rays from the inner 150 pc of the Galactic Center (GC). The main explanations proposed for this are: an unresolved population of millisecond pulsars (MSPs), dark matter (DM) annihilation,  and  cosmic rays interacting with molecular clouds at the GC. In this conference proceeding article we  highlight some of the cosmic-ray results
of \cite{MaciasGordon2014}. 
The MSPs and DM explanations were modeled with spatial templates well fitted by the square of a generalized Navarro-Frenk-White (NFW)  profile with inner slope $\gamma=1.2$.
The cosmic-ray option was modeled with a 20-cm continuum emission Galactic Ridge template. A template based on the HESS residuals were shown to give similar results.
The gamma-ray excess was found to be best fit by a combination of the generalized NFW squared template and a Galactic Ridge template. We also found the spectra of each template was not significantly affected in the combined fit and is consistent with previous single template fits. Due to the ridge like morphology of the molecular gas at the GC, the cosmic-ray option was not able to explain all the excess gamma rays, indicating that  DM annihilation and/or an unresolved population of MSPs is also needed.}

\FullConference{Cosmic Rays and the InterStellar Medium - CRISM 2014,\\
		24-27 June 2014\\
		Montpellier, France}

\begin{document}

\section{Introduction}
\label{sec:introduction}

Gamma rays constitute an excellent search channel for a signature of pair annihilation of dark matter (DM), since they can propagate almost without absorption from the source to the observer. Amongst all possible target regions in the gamma-ray sky, the Galactic Center (GC) is expected to be the brightest DM emitting source as it is relatively close by and has a high density of DM. However, this region is populated by a variety of astrophysical gamma-ray sources that make it hard to unambiguously identify a DM signal~\cite{cirelli2,Funk}. 

Several groups, using data from the Fermi large area  telescope (Fermi-LAT), have identified an
extended GeV source of $\gamma$-rays in the GC region
\cite{Vitale:2009hr,Hooper:2010mq,Morselli:2010ty,Boyarsky:2010dr,hooper,Abazajian:2012pn,hooperkelsoqueiroz2012,GordonMacias2013,MaciasGordon2014,Abazajian2014,Daylan:2014}. 
It has been confirmed that this GeV extended source is best fitted by the combination of a spherically-symmetric template 
and a ridge-like template resembling the central molecular zone (CMZ) in the inner $\sim 200$ pc of the GC~\cite{MaciasGordon2014}. For the spherical component, the square of a generalized Navarro-Frenk-White (NFW) profile with inner slope $\gamma = 1.2$ provides a good fit, and for the GeV Galactic Ridge component, a template based on the HESS residuals or 20-cm continuum radio emission map provides a satisfactory fit. The spectrum of each template is not significantly affected in the combined fit and is consistent with previous single-template fits. 
The statistical independence of these two GeV extended sources allows us to robustly extract spectral and spatial GeV $\gamma$-ray information from the Galactic Ridge.   
There have been three main proposals for the origin of the spherically symmetric emission: an unresolved population of millisecond pulsars (MSPs), dark matter (DM) pair annihilation, and cosmic rays (CRs) interacting with the interstellar medium and radiation field at the GC.  Some examples of recent articles discussing the pros and cons of the different proposals are~\cite{MaciasGordon2014,Abazajian2014,Daylan:2014,gomez2013,Lacroix:2014,Bringmann:2014lpa,Cirelli:2014lwa,cholis2014,CarlsonProfumo,
Petrovic:2014uda,Yuan:2014rca,Cholis:2014lta,Gomez-Vargas2014,Macias2014}. 

The main aim  this conference proceeding article is to  highlight some of the CR  results	
of \cite{MaciasGordon2014}. We also relate those results to earlier work done in \cite{GordonMacias2013} and some of the more recent findings of Ref.~\cite{Macias2014}.


%
\section{Data Reduction}
\label{sec:data}

The Fermi-LAT data selection is the same as described in~\cite{GordonMacias2013}. In summary, we analysed Pass-7\footnote{Preliminary checks have shown our results are not significantly changed if we instead use Pass-7 reprocessed  data.} data taken within a squared region of $7^{\circ}\times7^{\circ}$ centred on Sgr A$^{\star}$ in the first 45 months of observations over the period August 4, 2008$-$June 6, 2012. We used the standard data cuts and kept only the \texttt{SOURCE} class events which have a high probability of being photons of astrophysical origin. We also selected events between 200 MeV$-$100 GeV without making any distinction between \textit{Front} and \textit{Back} events. 
The spectra were obtained by maximizing the likelihood of source models using the binned \textit{pyLikelihood} library in the Fermi Science Tools~\cite{fermitools}. We followed the same fitting procedure adopted in Ref.~\cite{AK,AKerratum} which has been recommended to be more suitable for crowded regions like the GC. Unless otherwise stated, the models included all sources suggested in the 2FGL~\cite{2FGL} catalog plus the LAT standard diffuse galactic background (DGB) and extragalactic background models \texttt{gal$_{-}$2yearp7v6$_{-}$v0.fits}  and  \texttt{iso$_{-}$p7v6source.txt} respectively.

\section{Multi-Wavelength Results}
\label{sec:Models}
The High Energy Stereoscopic System (HESS) collaboration~\cite{Aharonian:2006au} reported the discovery of diffuse  TeV emission from the innermost part of the GC  region. These $\gamma$-rays are localized over a ridge like area defined by Galactic longitude $\abs{ l } < 0.8^{\circ}$ and latitude $\abs{ b}<0.3^{\circ}$. Morphological analysis of the data shows a close correlation between $\gamma$-ray emission and the CMZ. This is a strong indication that the $\gamma$-rays originate in CR interactions with the interstellar gas, and are thus proportional to the product of the density of CRs and that of the target material.

As mentioned in the previous section, diffuse, non-thermal ridge like emission is also seen by the Fermi large area  telescope (Fermi-LAT) in the GeV range. It is also seen
 by radio telescopes in the GHz range. 
 %
 A  CR flare, occurring in the GC,  $10^4$ years ago has been proposed to explain the TeV Galactic Ridge \cite{Aharonian:2006au}. 
 Alternative, steady-state models explaining all three data sets (TeV, GeV, and radio) have been proposed \cite{Yusef-Zadeh2013,Macias2014}. Ref.~\cite{Macias2014} showed that the flare models from the GC also provides an acceptable fit to the GeV and radio data.
However, it was found that
 the flare models did not fit the spherically symmetric GeV excess
 as well as the usual square of the generalized Navarro-Frenk-White spatial profile, but are better suited to explaining
  the Galactic Ridge gamma-ray excess.

\begin{figure}[p!]
\begin{center}

\begin{tabular}{ccc}
\centering
\includegraphics[width=0.33\linewidth]{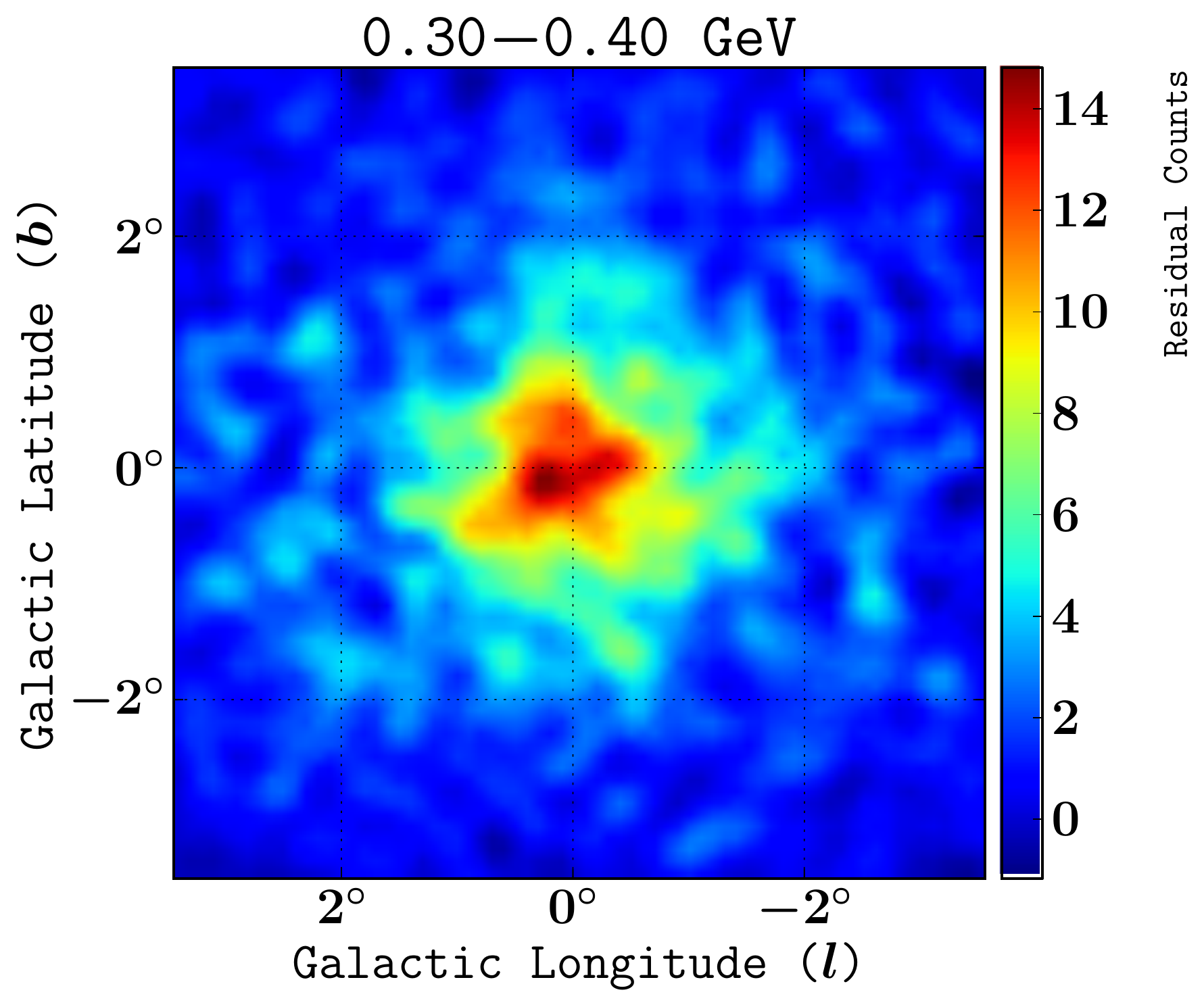} & \includegraphics[width=0.33\linewidth]{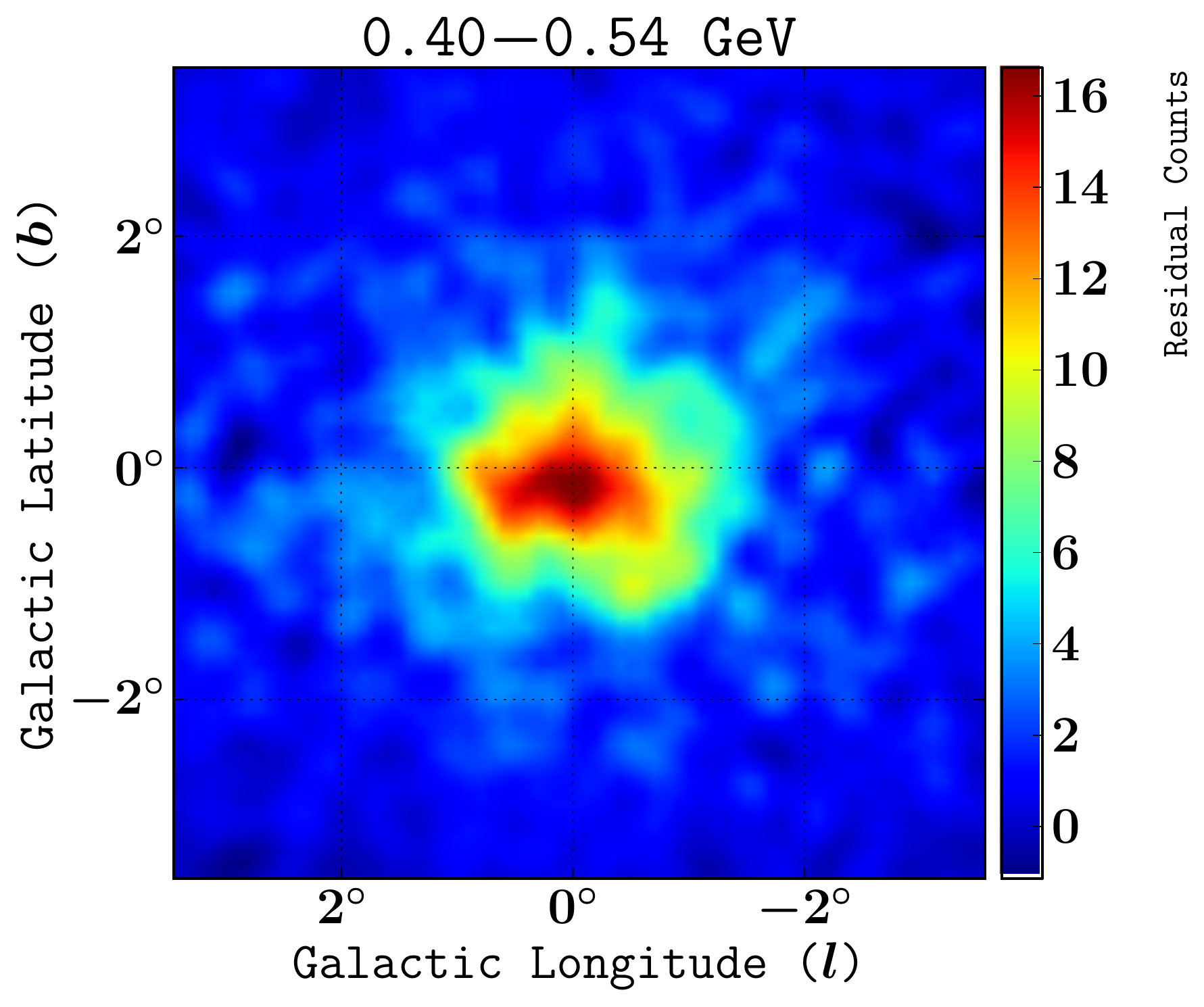} & \includegraphics[width=0.33\linewidth]{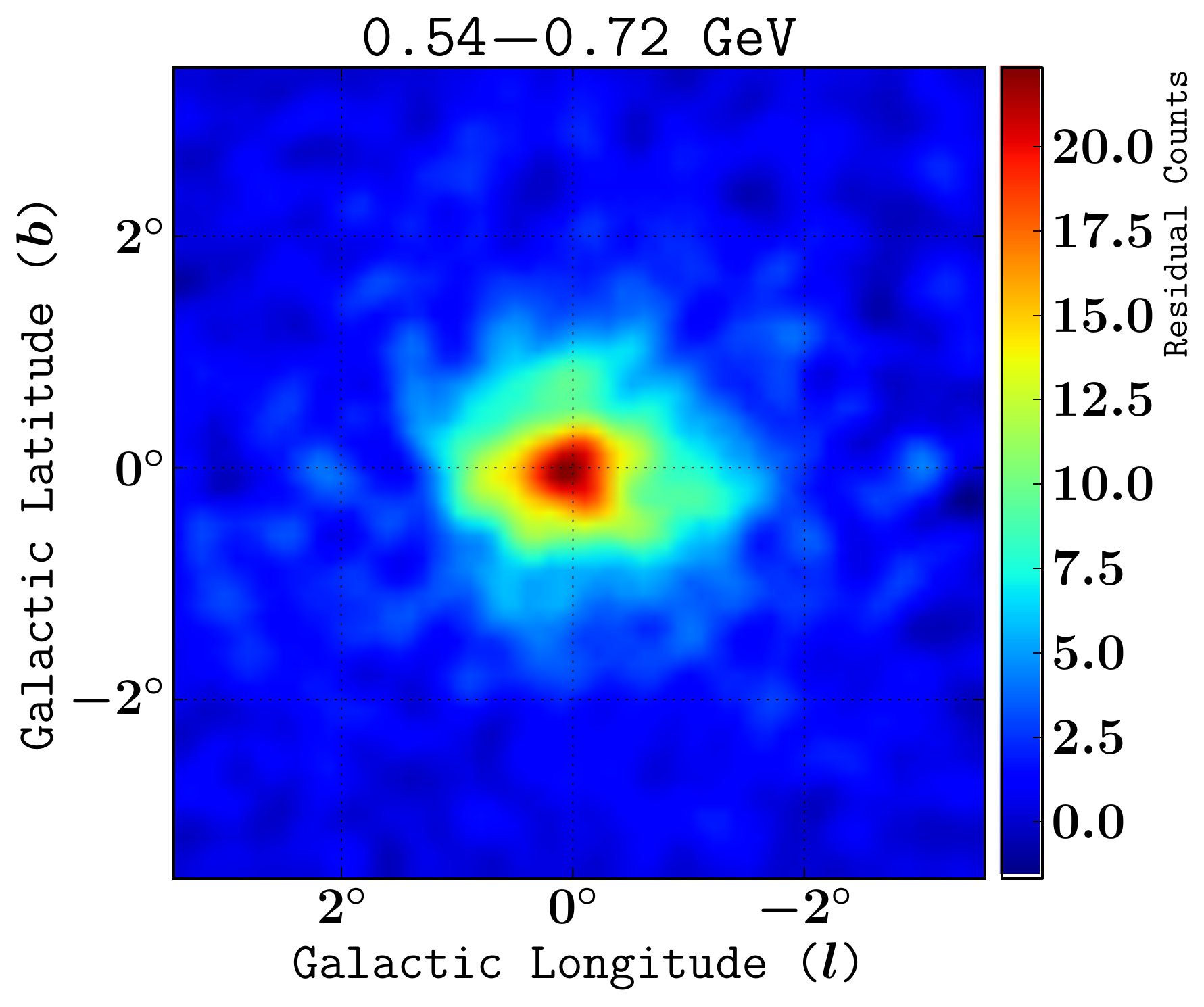}      

\end{tabular}

\begin{tabular}{ccc}
\centering
\includegraphics[width=0.33\linewidth]{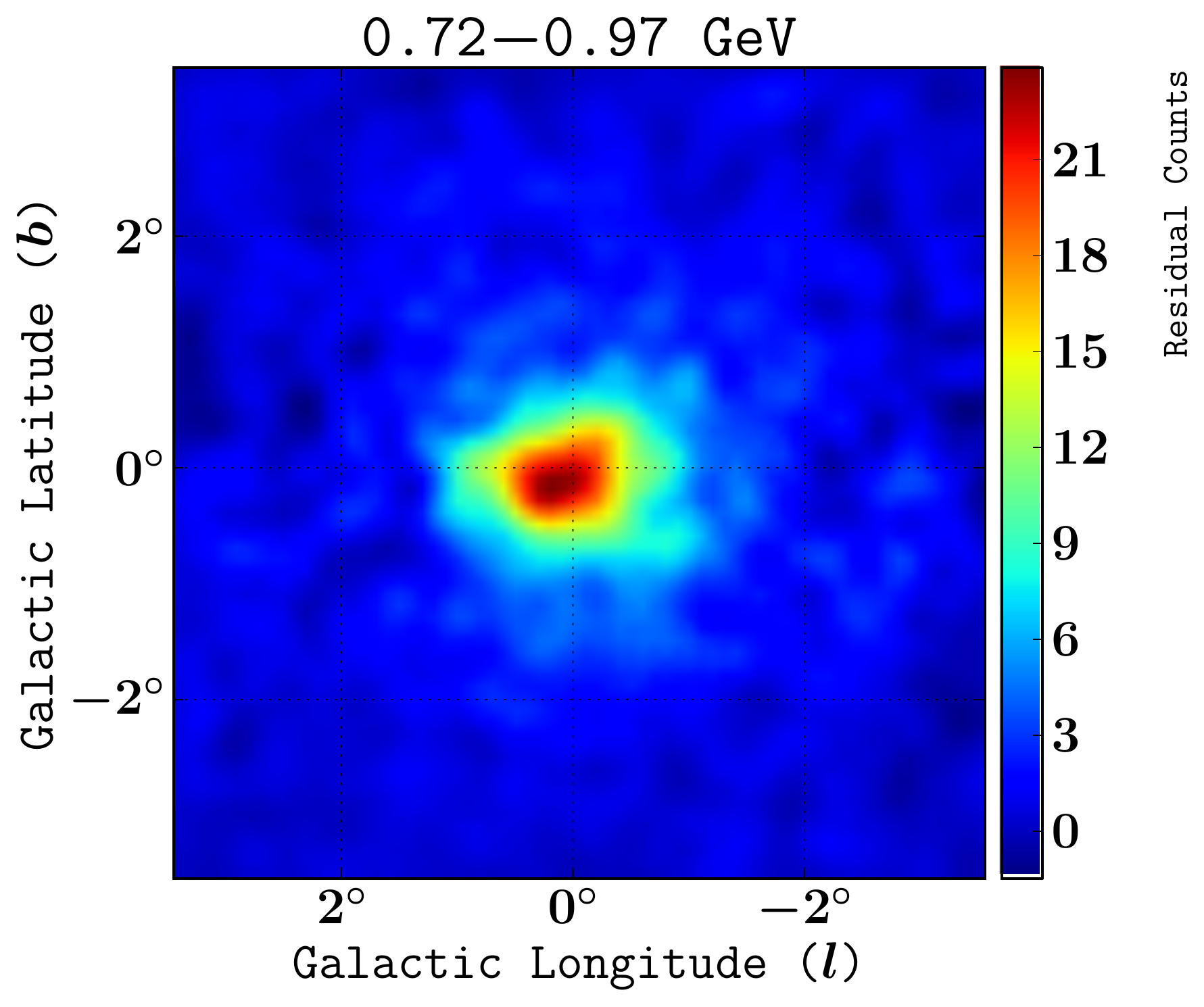} & \includegraphics[width=0.33\linewidth]{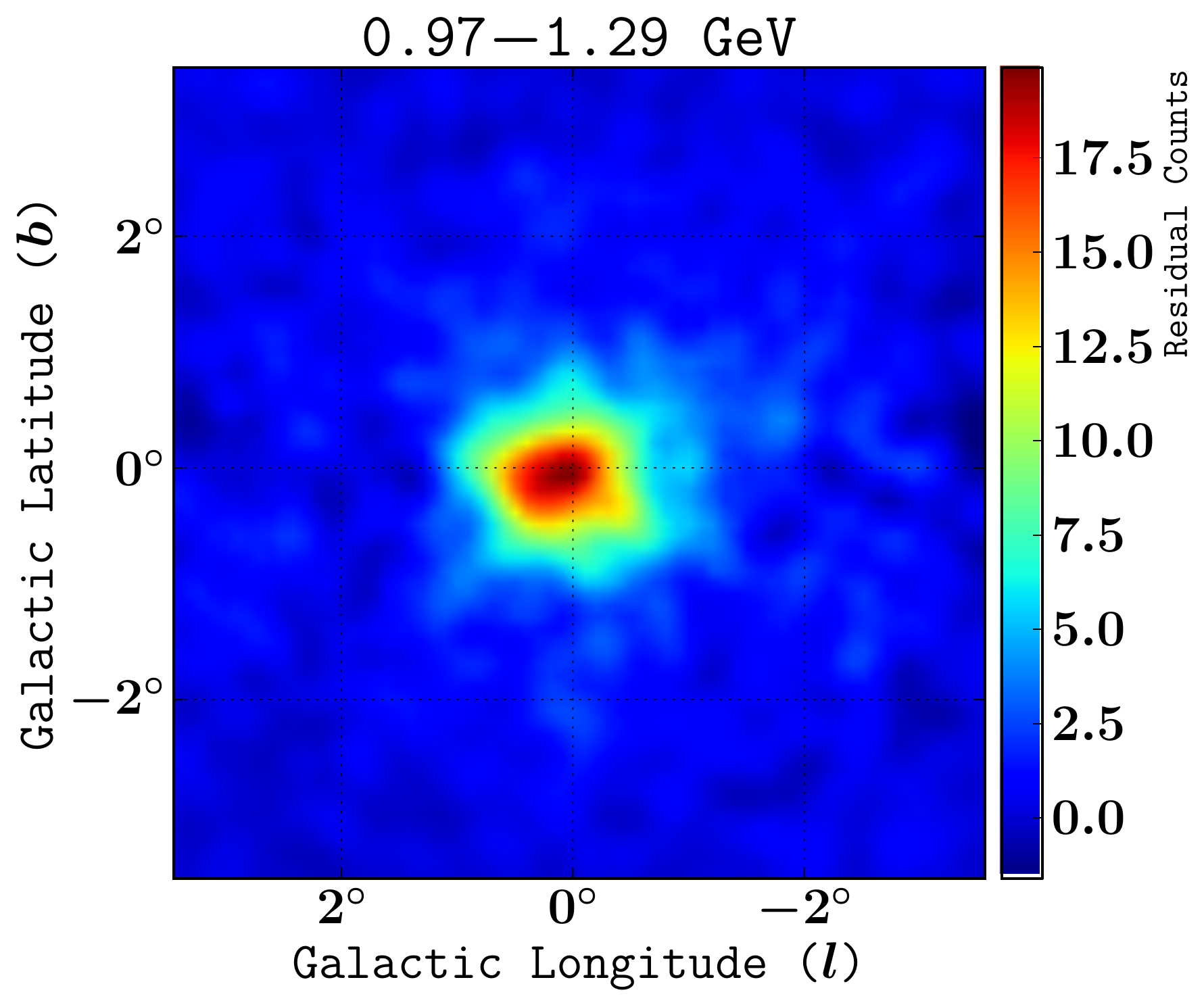} & \includegraphics[width=0.33\linewidth]{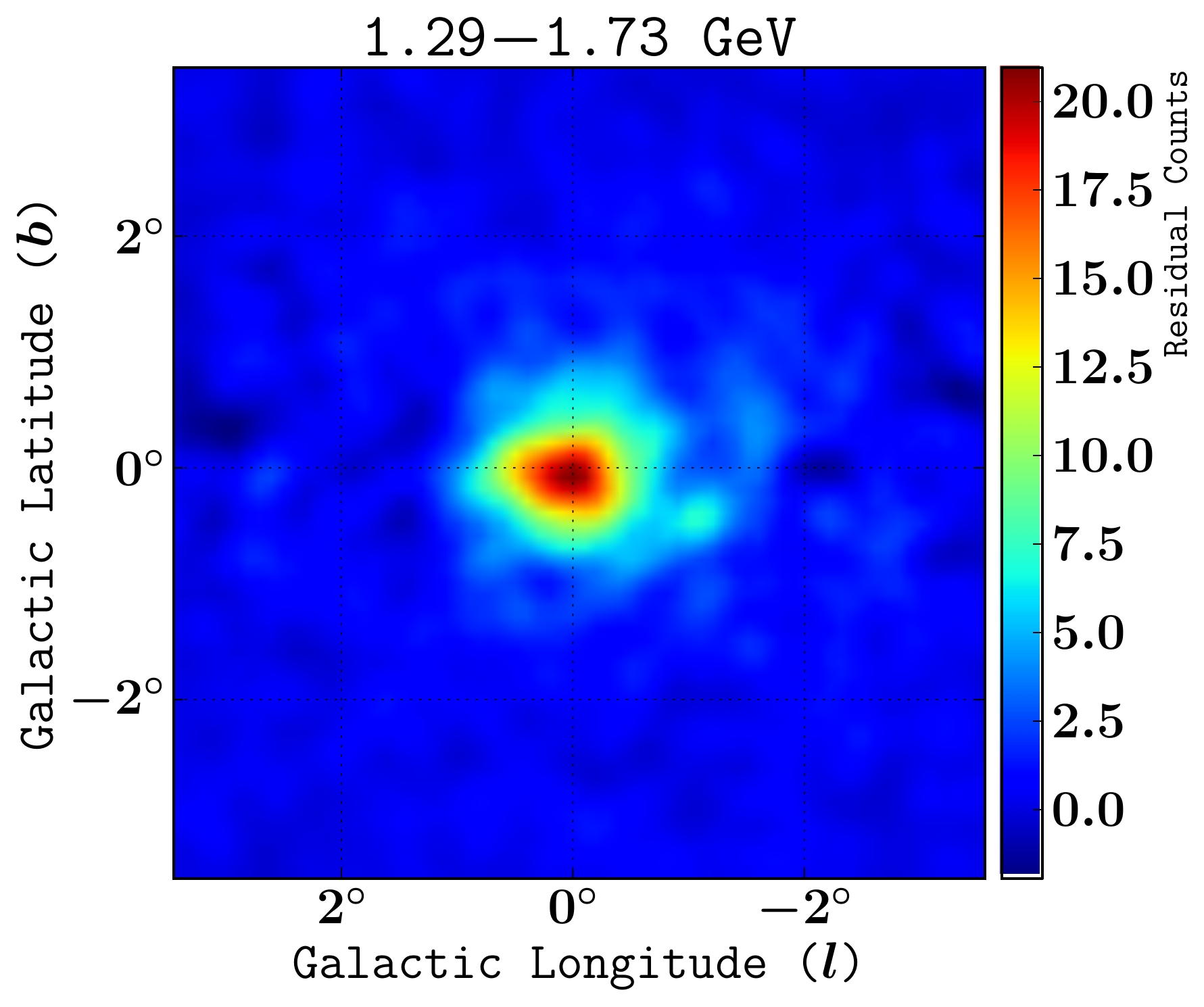}      

\end{tabular}

\begin{tabular}{ccc}
\centering
\includegraphics[width=0.33\linewidth]{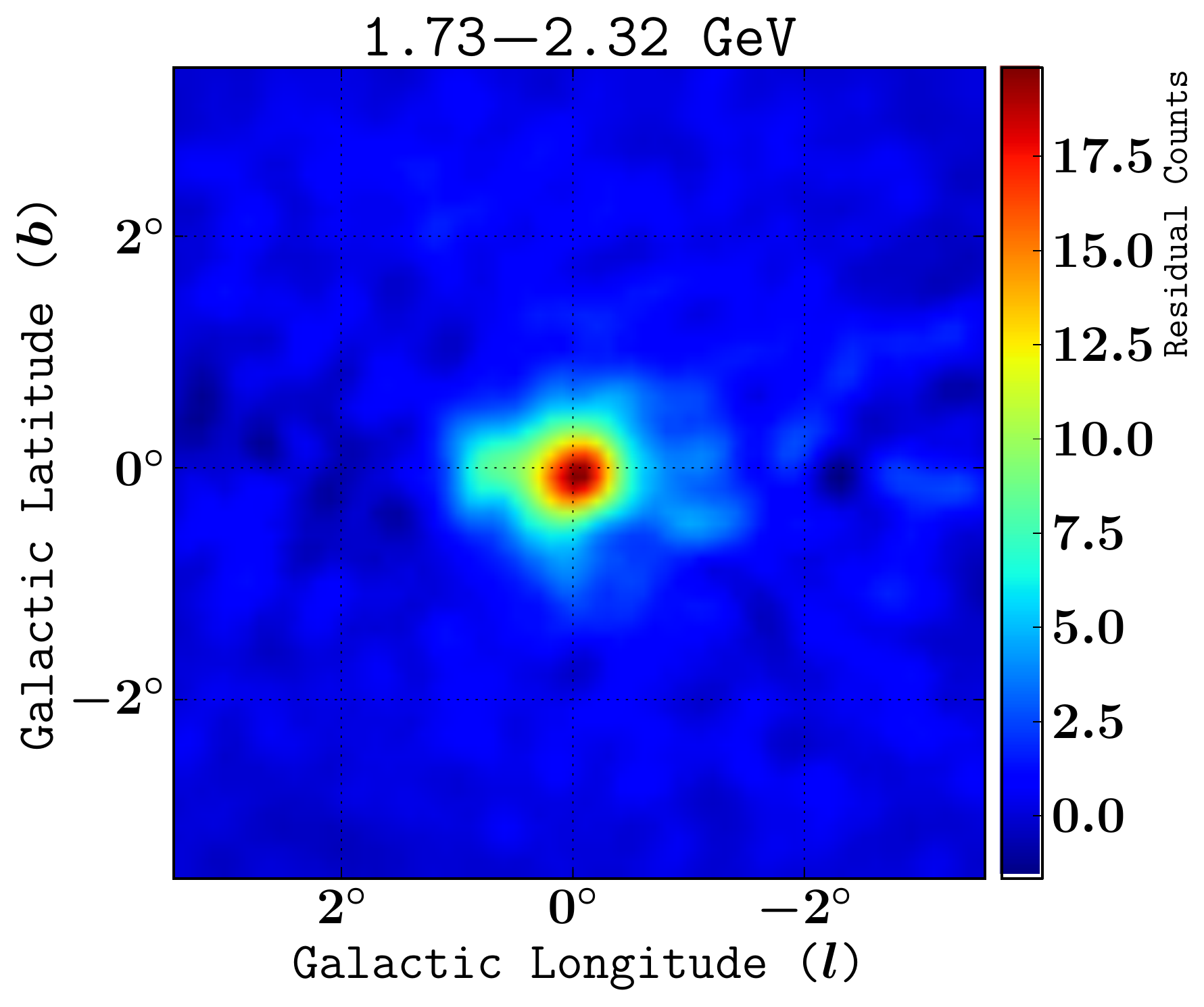} & \includegraphics[width=0.33\linewidth]{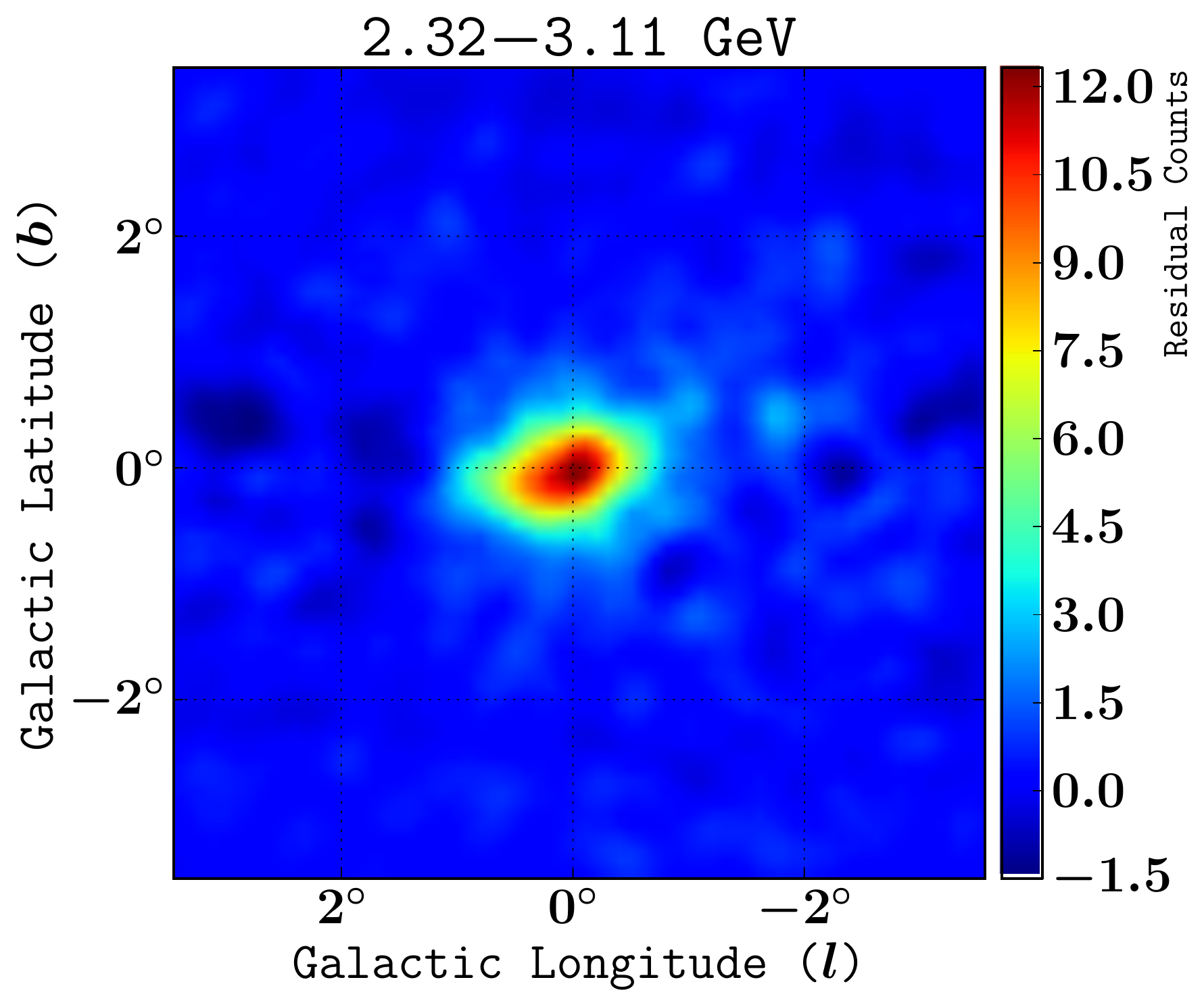} & \includegraphics[width=0.33\linewidth]{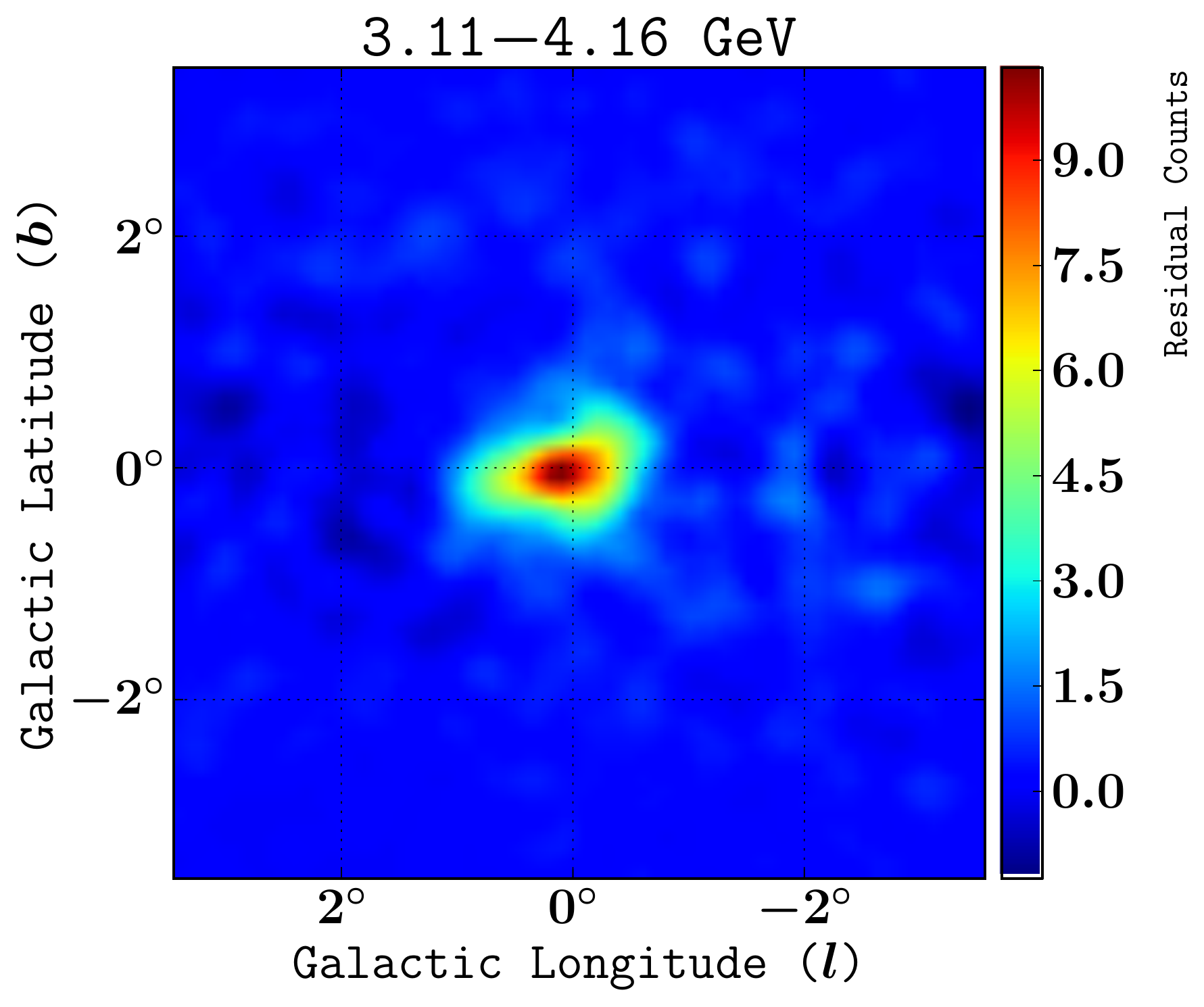}      

\end{tabular}

\begin{tabular}{lcc}

\includegraphics[width=0.33\linewidth]{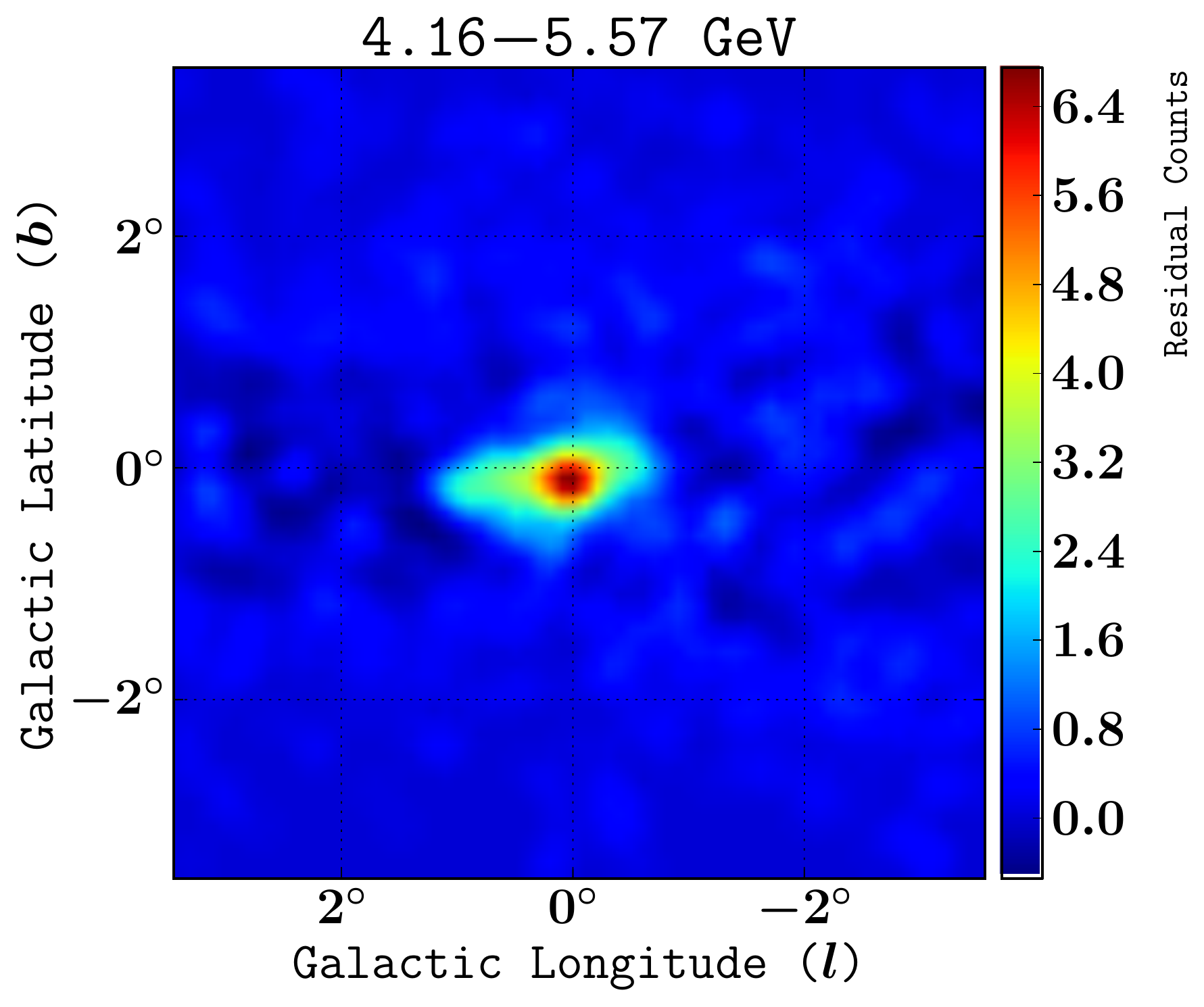} & \includegraphics[width=0.33\linewidth]{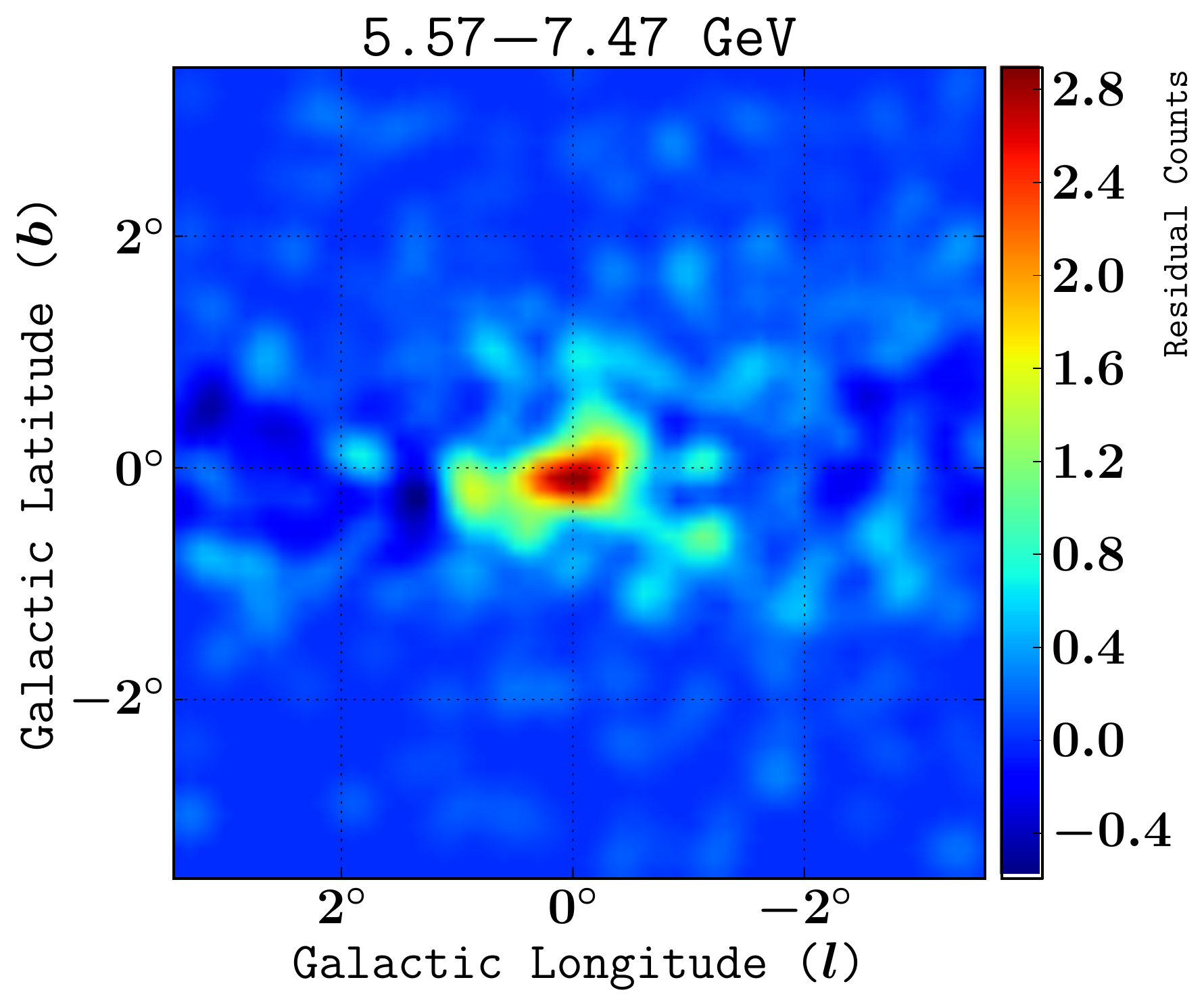}  & \includegraphics[width=0.33\linewidth]{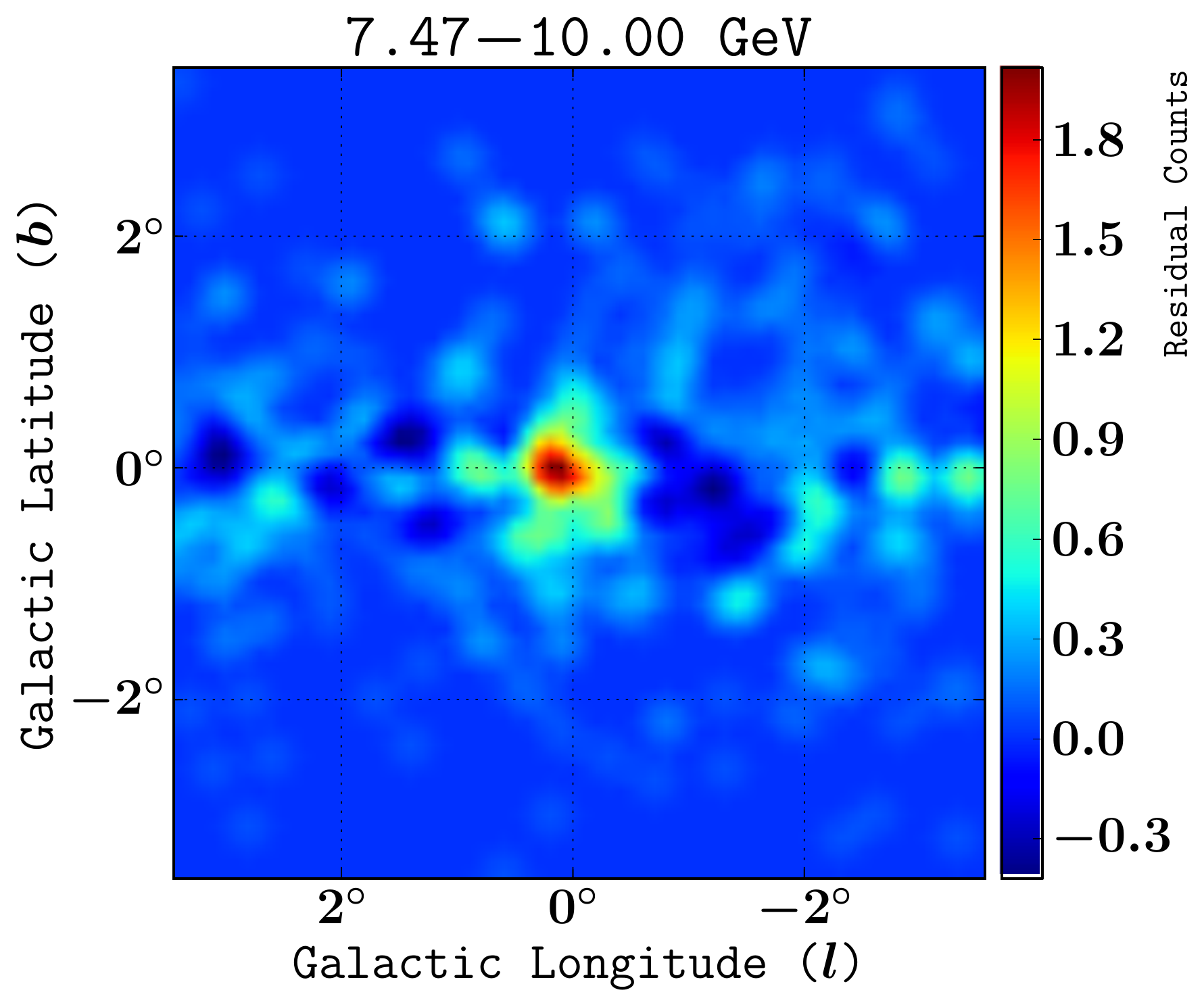}
\end{tabular}

\caption{ \label{fig:residuals} Shown are the best-fit model residuals where the spherical template and the 20-cm Galactic Ridge have not been subtracted from the data. The images have been smoothed with a 0.3$^\circ$ radius Gaussian filter.
}
\end{center}

\end{figure}

\begin{figure}[h!]
\begin{center}

\begin{tabular}{cc}

\includegraphics[width=0.5\linewidth]{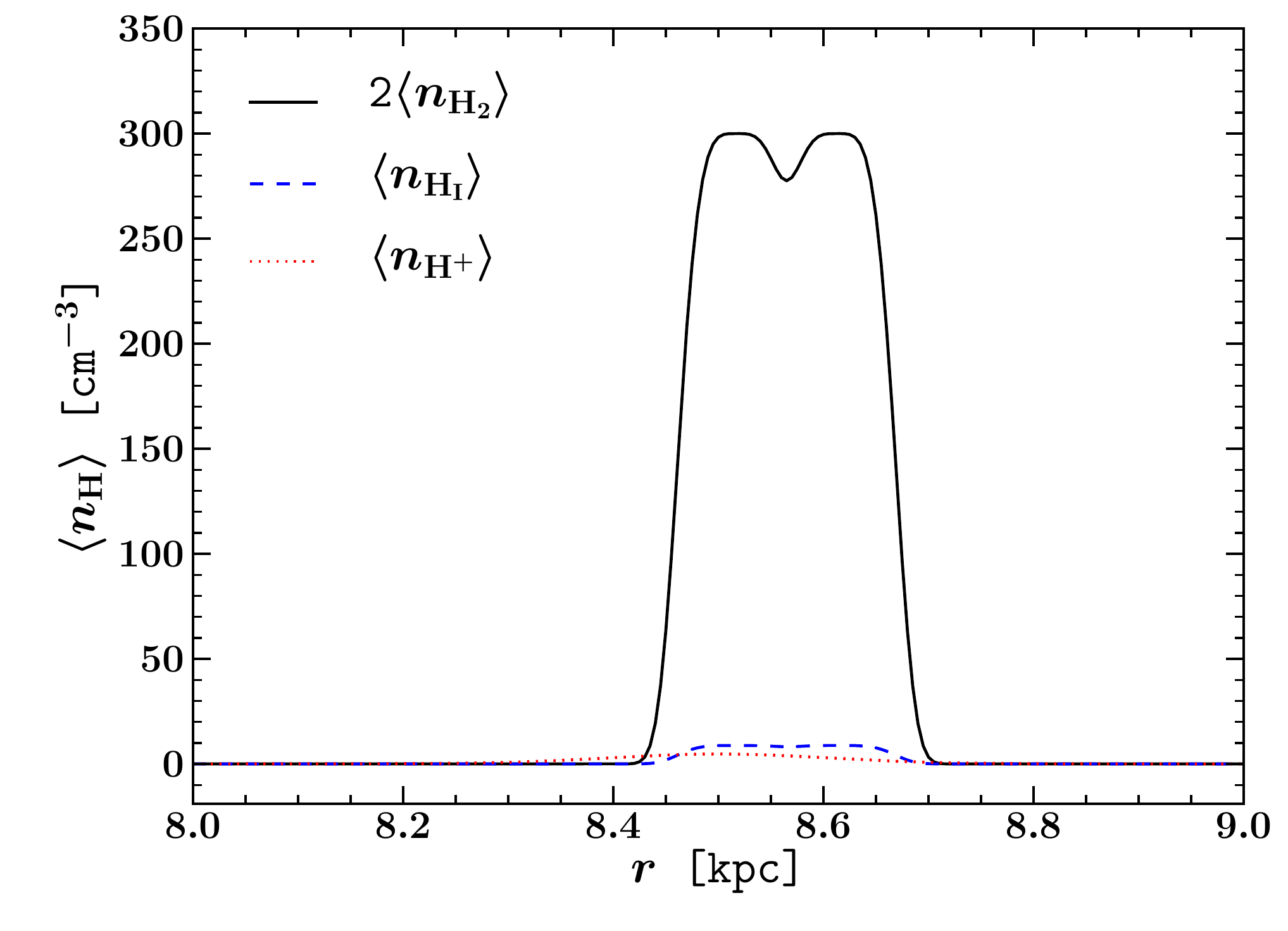} \includegraphics[width=0.5\linewidth]{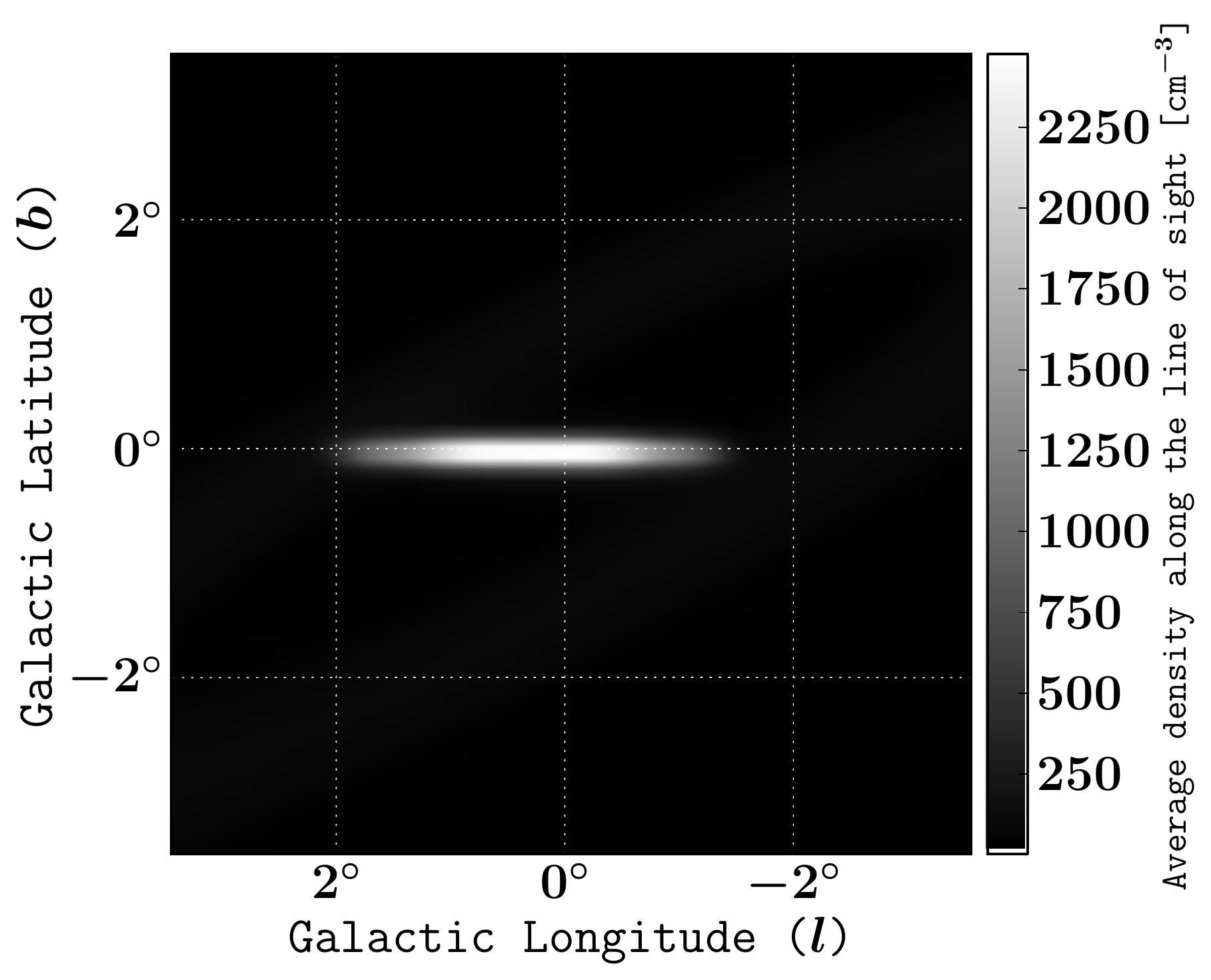}  
\end{tabular}
\centering
\includegraphics[width=0.5\linewidth]{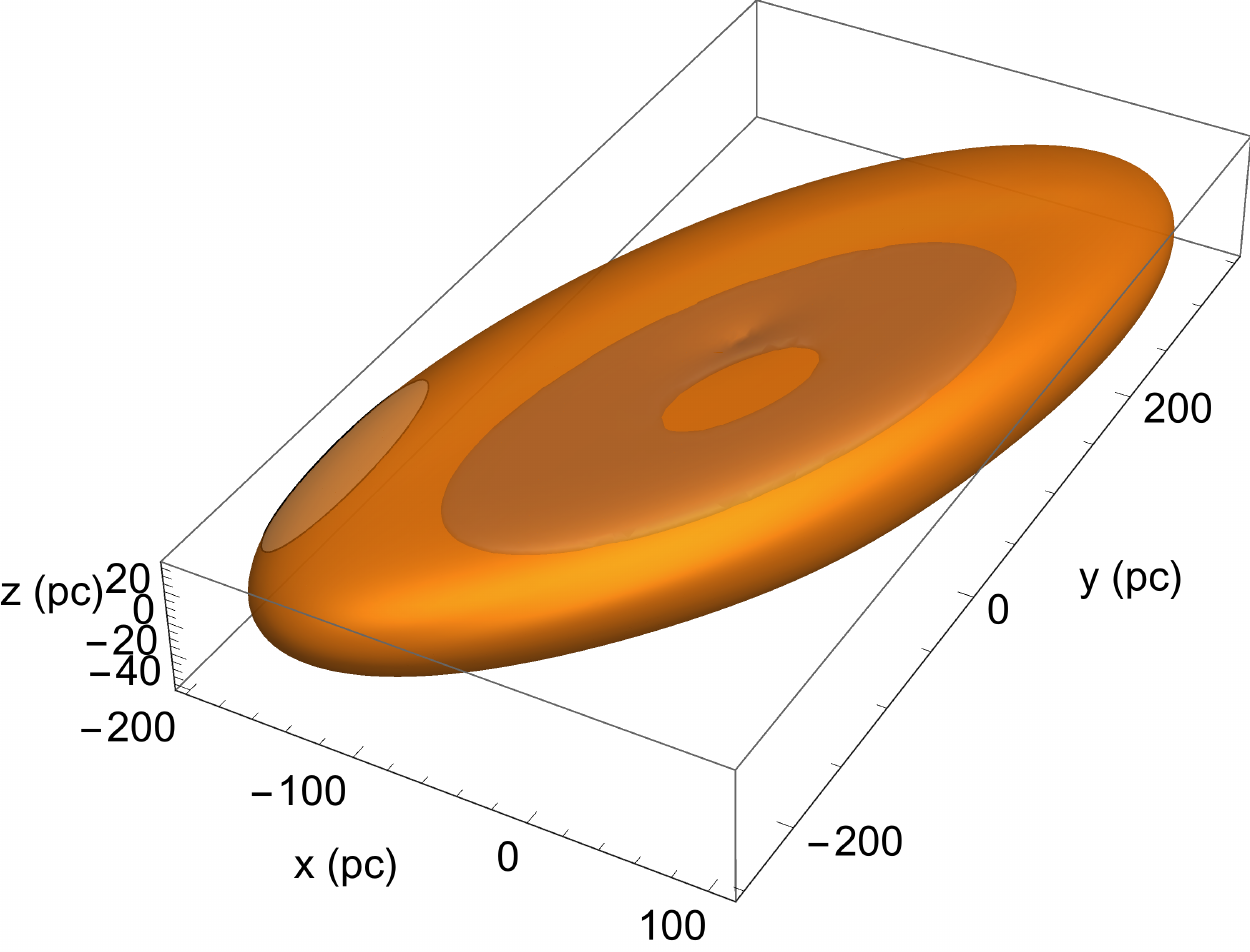} 

\caption{ \label{fig:disk}Panels shown are obtained from an interstellar gas model developed by Ref.~\cite{Ferriere}. \textit{Top left:} Densities of interstellar gas as functions of distance $(r)$ along the line of sight passing between us and the Galactic Center; Molecular hydrogen is shown with a black solid line, atomic hydrogen with a blue dashed line and ionized hydrogen atoms with a red dotted line. \textit{Top right:}  Total space-averaged density of interstellar hydrogen nuclei along the line of sight of the innermost 3 kpc.  \textit{Bottom:}  Three dimensional contour plot of total densities of interstellar gas. The yellow contour corresponds to 10 hydrogen nuclei per cm$^3$ and the blue one to 290  hydrogen nuclei per cm$^3$. The GC is at the origin and the Solar System is at  $x=8500$ pc, $y=0$ pc, $z=0$ pc.
}   
\end{center}
\end{figure}

\begin{figure}[h]
\begin{center}
\includegraphics[width=0.8\linewidth]{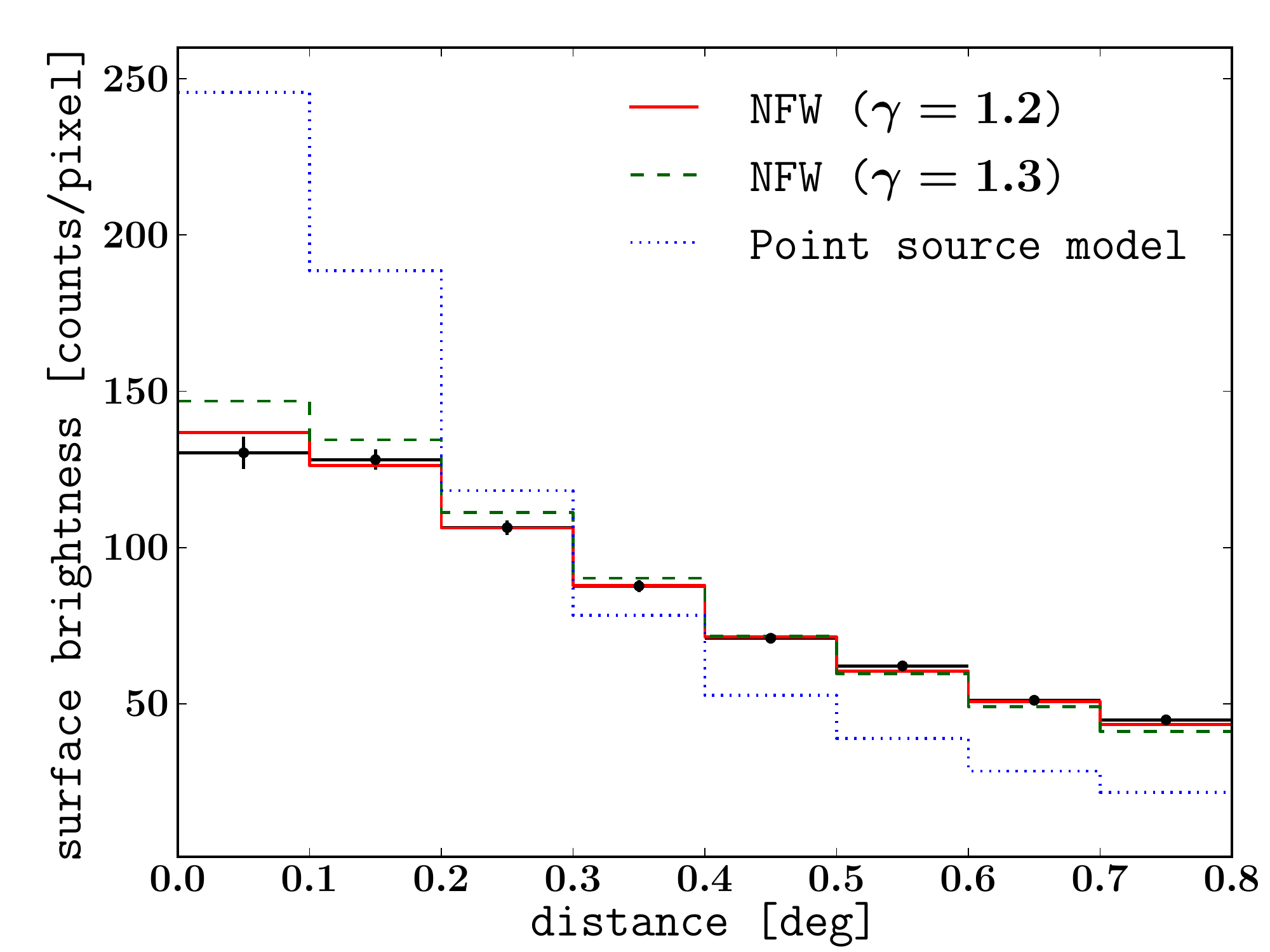} 
\end{center}
\caption{ \label{fig:profile}(a) Radial profile of the Fermi-LAT residuals when the spherical symmetric best-fit template is not subtracted
\cite{GordonMacias2013}. The plot is obtained from a ring analysis computed around Sgr A*. The histograms show the effective LAT point spread function (PSF) for three different profile models: (i) NFW with inner slope $\gamma\simeq 1.2$ (red continuous line) for which we get $\chi^2/\mbox{dof}=5.5/7$. (ii) NFW with $\gamma=1.3$ (green dashed line) and $\chi^2/\mbox{dof}=44.6/7$, and lastly (iii) the profile for a PS model (blue dotted line) with $\chi^2/\mbox{dof}=2479.9/7$. For all cases the spectra was modelled with a Log Parabola.}
\end{figure}

\section{Model Analysis}
\label{sec:Discussion}
In Ref~\cite{MaciasGordon2014}  the interaction of CRs with molecular clouds was evaluated as an explanation for the 
excess gamma-ray emission at the GC.
It was found that  the data prefer combinations of the Galactic Ridge template and a spherical template.


The  significance of adding a new component was checked by evaluating 
the test statistic (TS) which is defined as in Ref.~\cite{2FGL}:
\begin{eqnarray}\label{tsdef}
{\rm TS}=2\left[\log \mathcal{L} (\mbox{new source})-\log \mathcal{L} (\mbox{NO-new source})\right]\mbox{,} \qquad 
\end{eqnarray}
where $\mathcal{L}$ stands for the maximum of the likelihood of the data given the model with or without the new source.
In the large sample limit, under the no source hypothesis, TS has a $\chi^2/2$ distribution with the number of degrees of freedom equal to the number of 
parameters associated with the proposed positive amplitude new source  \cite{wilks,mattox}. As the amplitude is restricted to be non-negative, a $\chi^2/2$ distribution rather than the $\chi^2$ distribution is needed.

The improvement in the fit of the 20-cm Galactic Ridge relative to a model with only 2FGL sources was found to be TS$=648-425=213$ for $13-9=4$ extra degrees of freedom (dof). This corresponds to a $14\sigma$ detection (if we convert to the equivalent p-value for 1 degree of freedom) and so confirms that the 20-cm Galactic Ridge does improve the fit to the GC gamma-ray excess. However, the corresponding TS for the spherical template was found to be 870 and for only 3 extra dof and so clearly also improves the fit substantially. 

It was checked whether the 20-cm Galactic Ridge still improves the fit once the        spherical template is included. 
It was found that TS$=1330-1295=35$ for 4 extra dof which corresponds to a 5$\sigma$ detection. This showed that the GC gamma-ray excess motivated a sum of the spherical template and the Galactic Ridge being included.


The above analysis with the HESS residual Galactic Ridge and we found that a TS=30 for 4 extra dof  
which is less than the 20-cm case, but the difference is not statistically significant.
The need for the Galactic Ridge can be seen in the residuals shown in Fig.~\ref{fig:residuals}.


Additionally, Ref.~\cite{MaciasGordon2014} found that the inclusion of the Galactic Ridge does not significantly alter the spectral parameters of the spherical template.



%
Similar results were also found in Ref.~\cite{Macias2014} for flare models. There the emission of a flare  of CRs from the Sgr A* region 
produces a CR distribution with a radius which grows approximately proportional to the square-root of the flare age.
The disk-like shape of the spatial distribution of interstellar gas 
at the GC can be seen in Fig.~\ref{fig:disk}. The flare models can only amputate the edges of the disk by being of a sufficiently young age, but they cannot produce a sufficiently large spherical emission which drops as a generalized NFW squared with inner slope $\gamma=1.2$ as shown in the radial profile of Fig.~\ref{fig:profile}.

  \section{Conclusions}
Ref.~\cite{MaciasGordon2014} found that the GC gamma-ray excess is best fit by adding to the base 2FGL model both a 
spherically symmetric source  and a Galactic Ridge shaped source based on a 20-cm continuum emission template. Similar results were found for a Galactic Ridge template based on the HESS data residuals. The addition of the Galactic Ridge was not found to significantly affect the spherical source's spectral parameters.
Similar results were also found in Ref.~\cite{Macias2014} for flare models.  In summary, the spatial distribution of interstellar gas at the GC is too ridge like to explain  the spherical excess emission of gamma rays seen in the Fermi-LAT data.



\section*{Acknowledgements}
O.M. is supported by a UC Doctoral Scholarship. This work makes use of results obtained with \textsc{Fermi Science Tools}~\cite{fermitools}, \textsc{DMFit}~\cite{Profumo1}, \textsc{Minuit}~\cite{minuit},  \textsc{SciPy}~\cite{scipy}, and GALPROP~\cite{GALPROP, webrun}.


\bibliographystyle{hunsrt}
\bibliography{GalacticRidge_CRISM}

\end{document}